\documentclass[aps,showpacs,twocolumn,superscriptaddress]{revtex4}
\usepackage{amssymb}
\usepackage{epsfig}
\usepackage{graphicx,amsmath}
\begin{document}

\title{Behavior of the antiferromagnetic phase transition near the fermion
condensation quantum phase transition in $\rm \bf YbRh_2Si_2$}
\author{V.R. Shaginyan}\email{vrshag@thd.pnpi.spb.ru}
\affiliation{Petersburg Nuclear Physics Institute, RAS, Gatchina,
188300, Russia}\affiliation{Racah Institute of Physics, Hebrew
University, Jerusalem 91904, Israel}
\author{M.Ya. Amusia}\affiliation{Racah Institute
of Physics, Hebrew University, Jerusalem 91904, Israel}
\author{K.G. Popov}\affiliation{Komi Science Center, Ural Division,
RAS, Syktyvkar, 167982, Russia}

\begin{abstract}
Low-temperature specific-heat measurements on $\rm YbRh_2Si_2$ at
the second order antiferromagnetic (AF) phase transition reveal a
sharp peak at $T_N=$ 72 mK. The corresponding critical exponent
$\alpha$ turns out to be $\alpha=0.38$, which differs significantly
from that obtained within the framework of the fluctuation theory of
second order phase transitions based on the scale invariance, where
$\alpha\simeq0.1$. We show that under the application of magnetic
field the curve of the second order AF phase transitions passes into
a curve of the first order ones at the tricritical point leading to
a violation of the critical universality of the fluctuation theory.
This change of the phase transition is generated by the fermion
condensation quantum phase transition. Near the tricritical point
the Landau theory of second order phase transitions is applicable
and gives $\alpha\simeq1/2$. We demonstrate that this value of
$\alpha$ is in good agreement with the specific-heat measurements.
\end{abstract}

 \maketitle

\section{Introduction}

Fundamental understanding of the low-temperature physical
properties of such strongly correlated Fermi systems as heavy
fermion (HF) metals in the vicinity of a quantum phase transition
persists as one of the most challenging objectives of
condensed-matter physics. List of these extraordinary properties
are markedly large. Recent exciting measurements on $\rm
YbRh_2Si_2$ at the second order antiferromagnetic (AF) phase
transition extended the list and revealed a sharp peak in
low-temperature specific heat, which is characterized by the
critical exponent $\alpha=0.38$ and therefore differs drastically
from those of the conventional fluctuation theory of second order
phase transitions \cite{TNsteg}, where $\alpha\simeq0.1$
\cite{land1}. The obtained large value of $\alpha$ casts doubts on
the applicability of the conventional theory and sends a real
challenge for theories describing the second order phase
transitions in HF metals \cite{TNsteg}, igniting strong theoretical
effort to explain the violation of the critical universality in
terms of the tricritical point \cite{sap,misav,imada,kling}.

The striking feature of the fermion condensation quantum phase
transition (FCQPT) is that it has profound influence on
thermodynamically driven second order phase transitions provided
that these take place in the non-Fermi liquid (NFL) region formed
by FCQPT \cite{ams,obz}. As a result, the curve of any second order
phase transition passes into a curve of the first order one at the
tricritical point leading to a violation of the critical
universality of the fluctuation theory. For example, the second
order superconducting phase transition in $\rm CeCoIn_5$ changes to
the first one in the NFL region \cite{shag1}. As we shall see, it
is this feature that gives the key to resolve the challenge.

It is a common wisdom that low-temperature and quantum fluctuations
at quantum phase transitions form the specific heat, magnetization,
magnetoresistance etc., which are drastically different from that
of conventional metals \cite{senth,col,lohneysen,si,sach}. Usual
arguments that quasiparticles in strongly correlated Fermi liquids
"get heavy and die" at a quantum critical point commonly employ the
well-known formula basing on assumptions that the $z$-factor (the
quasiparticle weight in the single-particle state) vanishes at the
points of second-order phase transitions \cite{col1}. However, it
has been shown that this scenario is problematic \cite{khodz}. On
the other hand, facts collected on HF metals demonstrate that the
effective mass strongly depends on temperature $T$, doping (or the
number density) $x$, applied magnetic fields $B$ etc, while the
effective mass $M^*$ itself can reach very high values or even
diverge, see e.g. \cite{lohneysen,si}. Such a behavior is so
unusual that the traditional Landau quasiparticles paradigm does
not apply to it. The paradigm says that elementary excitations
determine the physics at low temperatures. These behave as Fermi
quasiparticles and have a certain effective mass $M^*$ which is
independent of $T$, $x$, and $B$ and is a parameter of the theory
\cite{land}. A concept of FCQPT preserving quasiparticles and
intimately related to the unlimited growth of $M^*$ had been
developed in Refs. \cite{khs,ams,volovik}. In contrast to the
Landau paradigm based on the assumption that $M^*$ is a constant,
the FCQPT approach supports an extended paradigm, the main point of
which is that the well-defined quasiparticles determine the
thermodynamic and transport properties of strongly correlated Fermi
systems, $M^*$ becomes a function of $T$, $x$, $B$, while the
dependence of the effective mass on $T$, $x$, $B$ gives rise to the
non-Fermi liquid behavior \cite{obz,khodb,zph,ckz,plaq}. Studies
show that the extended paradigm is capable to deliver an adequate
theoretical explanation of the NFL behavior in different HF metals
and HF systems \cite{obz,khodb,ckz,plaq,shag1,shag2,shag3}.

In the present short communication, we analyze the specific-heat
measurements on $\rm YbRh_2Si_2$ in the vicinity of the second
order AF phase transition with $T_N=72$ mK \cite{TNsteg}. The
measurements reveal that the corresponding critical exponent
$\alpha=$ 0.38 which differs drastically from that produced by the
fluctuation theory of second order phase transitions, where
$\alpha\simeq$ 0.1. We show that under the application of magnetic
field $B$ the curve $T_N(B)$ of the second order AF phase
transitions in  $\rm YbRh_2Si_2$ passes into a curve of the first
order ones at the tricritical point with temperature
$T_{cr}=T_N(B_{cr})$. This change is generated by FCQPT. Near the
tricritical point the Landau theory of second order phase
transitions is applicable and gives $\alpha\simeq1/2$ \cite{land1}.
This value of $\alpha$ is in good agreement with the specific-heat
measurements describing the data in the entire temperature range
around the AF phase transition. As a result, we conclude that the
critical universality of the fluctuation theory is violated at the
line of the AF phase transitions due to the tricritical point.

\section{Fermion condensation quantum phase transition}

We start with visualizing the main properties of FCQPT. To this
end, consider the density functional theory for superconductors
(SCDFT) \cite{gross}. SCDFT states that the thermodynamic potential
$\Phi$ is a universal functional of the number density $n({\bf r})$
and the anomalous density (or the order parameter) $\kappa({\bf
r},{\bf r}_1)$ and provides a variational principle to determine
the densities \cite{gross}. At the superconducting transition
temperature $T_c$ a superconducting state undergoes the second
order phase transition. Our goal now is to construct a quantum
phase transition which evolves from the superconducting one.

Let us assume that the coupling constant $\lambda$ of the BCS-like
pairing interaction vanishes, $\lambda\to0$, making vanish the
superconducting gap at any finite temperature. In that case,
$T_c\to0$ and the superconducting state takes place at $T=0$ while
at finite temperatures there is a normal state. This means that at
$T=0$ the anomalous density
\begin{equation}\label{ANOM}\kappa({\bf
r},{\bf r_1})=\langle\Psi\uparrow({\bf r})\Psi\downarrow({\bf
r_1})\rangle\end{equation} is finite, while the superconducting gap
\begin{equation}\label{DEL}\Delta({\bf
r})=\lambda\int\kappa({\bf r},{\bf r_1})d{\bf r_1}\end{equation} is
infinitely small \cite{obz,shag1}. In Eq. \eqref{ANOM}, the field
operator $\Psi_{\sigma}({\bf r})$ annihilates an electron of spin
$\sigma, \sigma=\uparrow,\downarrow$ at the position ${\bf r}$. For
the sake of simplicity, we consider a homogeneous electron liquid
\cite{obz}. Then at $T=0$, the thermodynamic potential $\Phi$
reduces to the ground state energy $E$ which turns out to be a
functional of the occupation number $n({\bf p})$ since the order
parameter $\kappa({\bf p})=\sqrt{n({\bf p})(1-n({\bf p}))}$
\cite{dft,gross,yakov,plaq}. Upon minimizing $E$ with respect to
$n({\bf p})$, we obtain \cite{obz,khs}
\begin{equation}\label{FCM}
\frac{\delta E}{\delta n({\bf p})}=\varepsilon({\bf
p})=\mu,\end{equation} where $\mu$ is the chemical potential. As
soon as Eq. \eqref{FCM} has nontrivial solution $n_0({\bf p})$ then
instead of the Fermi step, we have $0<n_0({\bf p})<1$ in certain
range of momenta $p_i\leq p\leq p_f$ with $\kappa({\bf
p})=\sqrt{n_0({\bf p})(1-n_0({\bf p}))}$ is finite in this range,
while the single particle spectrum $\varepsilon({\bf p})$ is flat.
Thus, the step-like Fermi filling inevitably undergoes
restructuring and forms the fermion condensate (FC) when Eq.
\eqref{FCM} possesses for the first time the nontrivial solution at
some quantum critical point (QCP) $x=x_c$. Here $p_F$ is the Fermi
momentum and $x =p_F^3/3\pi^2$. In that case, the range vanishes,
$p_i\to p_f\to p_F$, and the effective mass $M^*$ diverges at QCP
\cite{khs,obz,ckz,khodb}
\begin{equation}\label{EFM}
\frac{1}{M^*(x\to x_c)}=\frac{1}{p_F}\frac{\partial\varepsilon({\bf
p})}{\partial{\bf p}}|_{p\to p_F;\,x\to x_c}\to 0.\end{equation} At
any small but finite temperature the anomalous density $\kappa$ (or
the order parameter) decays and, as we shall see, this state
undergoes the first order phase transition and converts into a
normal state characterized by the thermodynamic potential $\Phi_0$.
At $T\to0$, the entropy $S=-\partial \Phi_0/\partial T$ of the
normal state is given by the well-known relation \cite{land}
\begin{eqnarray}
S_0[n_0({\bf p})]&=& -2\int[n_0({\bf p})
\ln (n_0({\bf p}))+(1-n_0({\bf p}))\nonumber\\
&\times&\ln(1-n_0({\bf p}))]\frac{d{\bf p}}{(2\pi)^3},\label{SN}
\end{eqnarray}
which follows from combinatorial reasoning. It is seen from Eq.
\eqref{SN} that the normal state is characterized by the
temperature-independent entropy $S_0$ \cite{obz,yakov}. Since the
entropy of the superconducting ground state is zero, we conclude
that the entropy is discontinuous at the phase transition point,
with its discontinuity $\Delta S=S_0$. Thus, the system undergoes
the first order phase transition. The heat $q$ of transition from
the asymmetrical to the symmetrical phase is $q=T_cS_0=0$ since
$T_c=0$. Because of the stability condition at the point of the
first order phase transition, we have $\Phi_0[n({\bf
p})]=\Phi[\kappa({\bf p})]$. Obviously the condition is satisfied
since $q=0$.

At $T=0$, a quantum phase transition is driven by a nonthermal
control parameter, e.g. the number density $x$. To clarify the role
of $x$, consider the effective mass $M^*$ which is related to the
bare electron mass $M$ by the well-known Landau Eq. \cite{land}
\begin{equation}\label{LANDM}
\frac{1}{M^*}=\frac{1}{M}+\int \frac{{\bf p}_F{\bf p_1}}{p_F^3}
F({\bf p_F},{\bf p}_1)\frac{\partial n(p_1,T)}{\partial p_1}
\frac{d{\bf p}_1}{(2\pi)^3}.
\end{equation}
Here we omit the spin indices for simplicity, $n({\bf p},T)$ is
quasiparticle occupation number, and $F$ is the Landau amplitude.
At $T=0$, Eq. \eqref{LANDM} reads \cite{pfit,pfit1}
\begin{equation}\label{MM*}
\frac{M^*}{M}=\frac{1}{1-N_0F^1(x)/3}.\end{equation} Here $N_0$ is
the density of states of a free electron gas and $F^1(x)$ is the
$p$-wave component of Landau interaction amplitude $F$. When at
some critical point $x=x_c$, $F^1(x)$ achieves certain threshold
value, the denominator in Eq. \eqref{MM*} tends to zero so that the
effective mass diverges at $T=0$ \cite{pfit,pfit1}. It follows from
Eq. \eqref{MM*} that beyond the critical quantum point $x_c$, the
effective mass becomes negative. To avoid unstable and physically
meaningless state with a negative effective mass and in accordance
with Eq. \eqref{EFM}, the system must undergo a quantum phase
transition at QCP with $x=x_c$, which is QCP of FCQPT
\cite{khs,ams,obz,khodb}.

\begin{figure} [! ht]
\begin{center}
\vspace*{-0.5cm}
\includegraphics [width=0.49\textwidth]{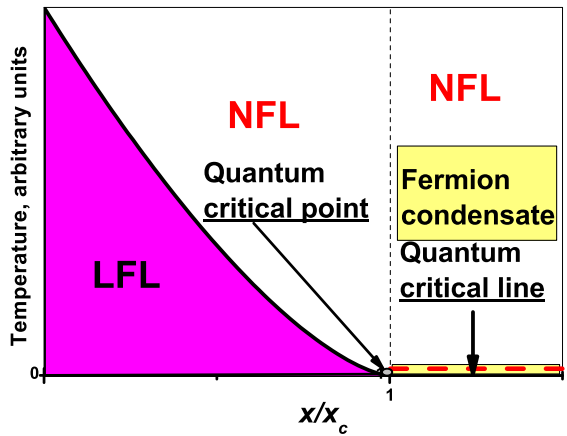}
\end{center}
\vspace*{-0.8cm} \caption{Schematic phase diagram of the system
driven to the FC state. The number density $x$ is taken as the
control parameter and depicted as $x/x_c$. The quantum critical
point, $x/x_c=1$, of FCQPT is shown by the arrow. At $x/x_c<1$ and
sufficiently low temperatures, the system in the Landau Fermi
liquid (LFL) state is shown by the shadow area. At $T=0$ and beyond
the critical point, $x/x_c>1$, the system is at the quantum
critical line depicted by the dash line and shown by the vertical
arrow. The critical line is characterized by the FC state with
finite superconducting order parameter $\kappa$. At $T_c=0$,
$\kappa$ is destroyed, the system undergoes the first order phase
transition and exhibits the NFL behavior at $T>0$.}\label{fig1}
\end{figure}

Schematic phase diagram of the system which is driven to FC by
variation of $x$ is reported in Fig. \ref{fig1}.  Upon approaching
the critical density $x_c$ the system remains in the Landau Fermi
liquid (LFL) region at sufficiently low temperatures
\cite{khodb,obz}, that is shown by the shadow area. At QCP $x_c$
shown by the arrow in Fig. \ref{fig1}, the system demonstrates the
NFL behavior down to the lowest temperatures. Beyond the critical
point at finite temperatures the behavior is remaining the NFL one
and is determined by the temperature-independent entropy $S_0$
\cite{obz,yakov}. In that case at $T\to 0$, the system is
approaching a quantum critical line (shown by the vertical arrow
and the dashed line in Fig. \ref{fig1}) rather than a quantum
critical point. Upon reaching the quantum critical line from the
above at $T\to0$ the system undergoes the first order quantum phase
transition, which is FCQPT taking place at $T_c=0$.

At $T>0$ the NFL state above the critical line, see Fig.
\ref{fig1}, is strongly degenerated, therefore it is captured by
the other states such as superconducting (for example, by the
superconducting state in $\rm CeCoIn_5$ \cite{shag1,shag2,yakov})
or by AF state (e.g. AF one in $\rm YbRh_2Si_2$ \cite{plaq})
lifting the degeneracy. The application of magnetic field
$B>B_{c0}$ restores the LFL behavior, where $B_{c0}$ is a critical
magnetic field, such that at $B>B_{c0}$ the system is driven
towards its LFL state \cite{obz,ckz,shag2}. In some cases, for
example in HF metal $\rm CeRu_2Si_2$, $B_{c0}=0$, see e.g.
\cite{takah}, while in $\rm YbRh_2Si_2$, $B_{c0}\simeq 0.06$ T
\cite{geg}. In our simple model of homogeneous electron liquid
$B_{c0}$ is taken as a parameter.

\section{${\bf T-B}$ phase diagram for $\rm \bf YbRh_2Si_2$ versus one for $\rm
\bf CeCoIn_5$}

\begin{figure}[!ht]
\begin{center}
\includegraphics [width=0.44\textwidth]{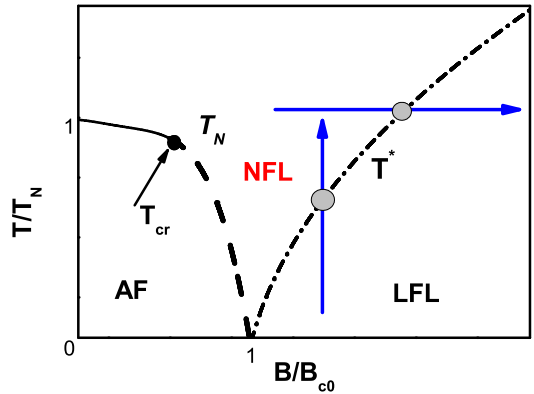}
\vspace*{-0.5cm}
\end{center}
\caption{ Schematic $T-B$ phase diagram for $\rm YbRh_2Si_2$. The
solid and dash $T_N(B)$ curves separate AF and non-Fermi liquid
(NFL) states representing the field dependence of the N\'eel
temperature. The black dot at $T=T_{cr}$ shown by the arrow  in the
dash curve is the tricritical point, at which the curve of second
order AF phase transitions shown by the solid line passes into the
curve of the first ones. At $T<T_{cr}$, the dash line represents
the field dependence of the N\'eel temperature when the AF phase
transition is of the first order. The NFL state is characterized by
the entropy $S_{0}$ given by Eq. \eqref{SN}. The dash-dot line
separating the NFL state and the weakly polarized LFL is
represented by $T^*(B/B_{c0})$ given by Eq. \eqref{BC0}. The
horizontal solid arrow represents the direction along which the
system transits from the NFL behavior to the LFL one at elevated
magnetic field and fixed temperature. The vertical solid arrow
represents the direction along which the system transits from the
LFL behavior to the NFL one at elevated temperature and fixed
magnetic field. The shadowed circle depict the transition
temperature $T^*$ from the NFL to LFL behavior.}\label{fig2}
\end{figure}

In Fig. \ref{fig2}, we present temperature $T/T_N$ versus field
$B/B_{c0}$ schematic phase diagram for $\rm YbRh_2Si_2$. There
$T_N(B)$ is the N\'eel temperature as a function of the magnetic
field $B$. The solid and dash lines indicate boundary of the AF
phase at $B/B_{c0}\leq 1$ \cite{geg}. For $B/B_{c0}\geq 1$, the
dash-dot line marks the upper limit of the observed LFL behavior.
This dash-dot line separates the NFL state and the weakly polarized
LFL, and is represented by  \cite{obz}
\begin{equation}\label{BC0}
\frac{T^*}{T_N}=a_1\sqrt{\frac{B}{B_{c0}}-1},
\end{equation}
where $a_1$ is a parameter. We note that Eq. \eqref{BC0} is in good
agreement with facts \cite{geg}. Thus, $\rm YbRh_2Si_2$
demonstrates two different LFL states, where the
temperature-dependent electrical resistivity $\Delta\rho$ follows
the LFL behavior $\Delta\rho\propto T^2$, one being weakly AF
ordered ($B\leq B_{c0}$ and $T<T_N(B)$) and the other being weakly
polarized ($B\geq B_{c0}$ and $T<T^*(B)$) \cite{geg}.
At elevated temperatures and fixed magnetic field the NFL state
occurs which is separated from the AF phase by the curve $T_N(B)$
of phase transition. In accordance with experimental facts we
assume that at relatively high temperatures $T/T_{N}(B)\simeq 1$
the AF phase transition is of the second order \cite{TNsteg,geg}.
In that case, the entropy and the other thermodynamic functions are
continuous functions at the curve of the phase transitions
$T_N(B)$. This means that the entropy of the AF phase $S_{AF}(T)$
coincides with the entropy $S(T)$ of the NFL state
\begin{equation} S_{AF}(T\to T_N(B))=S(T\to
T_N(B)).\label{TN}\end{equation} Since the AF phase demonstrates
the LFL behavior, that is $S_{AF}(T\to 0)\to0$, Eq. \eqref{TN}
cannot be satisfied at diminishing temperatures $T\leq T_{cr}$ due
to the temperature-independent term $S_0$ given by Eq. \eqref{SN}.
Thus, in the NFL region formed by FCQPT the second order AF phase
transition inevitably becomes the first order one at the
tricritical point with $T=T_{cr}$, as it is shown in Fig.
\ref{fig2}. At $T=0$, the critical field $B_{c0}$ is determined by
the condition that the ground state energy of the AF phase
coincides with the ground state energy of the weakly polarized LFL,
and the ground state of $\rm YbRh_2Si_2$ becomes degenerated at
$B=B_{c0}$. Therefore, the N\'eel temperature $T_N(B\to B_{c0})\to
0$.

\begin{figure} [! ht]
\begin{center}
\vspace*{-0.5cm}
\includegraphics [width=0.44\textwidth]{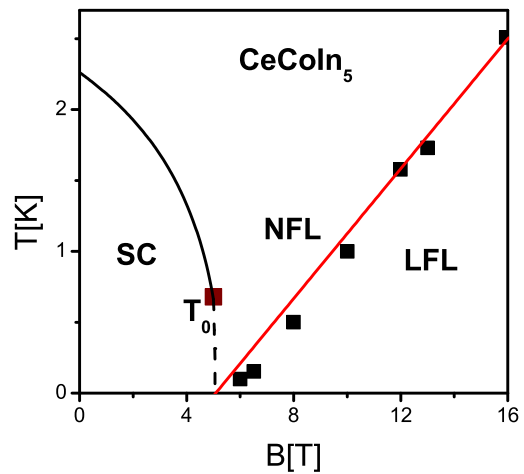}
\end{center}
\vspace*{-0.7cm} \caption{$B-T$ phase diagram of the $\rm CeCoIn_5$
heavy fermion metal. The interface between the superconducting and
normal phases is shown by the solid and dash lines. At $T<T_0$, the
curve of the second order superconducting phase transitions passes
into a curve of the first order ones at the tricritical point shown
by the square \cite{bian}. The interface between the
superconducting and normal phases is shown by the dashed line. The
solid straight line with the experimental points \cite{pag1} shown
by squares is the interface between the Landau Fermi liquid (LFL)
and non-Fermi liquid (NFL) states
\cite{shag1,bian,pag1}.}\label{CeCo}
\end{figure}

Upon comparing the phase diagram of $\rm YbRh_2Si_2$  depicted in
Fig.  \ref{fig2} with that of $\rm CeCoIn_5$ shown in Fig.
\ref{CeCo}, it is possible to conclude that they are similar in
many respects. Indeed, the line of the second order superconducting
phase transitions changes to the line of the first ones at the
tricritical point shown by the the square in Fig. \ref{CeCo}. This
transition takes place under the application of magnetic field $B>
B_{c2}\geq B_{c0}$ \cite{shag1,shag2}, where $B_{c2}$ is the
critical field destroying the superconducting state, and $B_{c0}$
is the critical field at which the magnetic field induced QCP takes
place \cite{bian,pag1}. We note that the superconducting boundary
line $B_{c2}(T)$ at lowering temperatures acquires the tricritical
point due to Eq. \eqref{TN} that cannot be satisfied at diminishing
temperatures $T\leq T_{cr}$, i.e. the corresponding phase
transition becomes first order \cite{shag1,shag2,bian}. This
permits us to conclude that at lowering temperatures, in the NFL
region formed by FCQPT the curve of any second order phase
transition passes into the curve of the first order one at the
tricritical point.

\section{The tricritical point in the ${\bf B-T}$ phase diagram of $\rm \bf YbRh_2Si_2$}

The Landau theory of the second order phase transitions is
applicable as the tricritical point is approached, $T\simeq
T_{cr}$, since the fluctuation theory can lead only to further
logarithmic corrections to the values of the critical indices.
Moreover, near the tricritical point, the difference
$T_N(B)-T_{cr}$ is a second order small quantity when entering the
term defining the divergence of the specific heat \cite{land1}. As
a result, upon using the Landau theory we obtain that the
Sommerfeld coefficient $\gamma_0=C/T$ varies as $\gamma_0\propto
|t-1|^{-\alpha}$ where $t=T/T_N(B)$ with the exponent is
$\alpha\simeq0.5$ as the tricritical point is approached at fixed
magnetic field \cite{land1}. We will see that $\alpha=0.5$ gives
good description of the facts collected in measurements of the
specific heat on $\rm YbRh_2Si_2$. Taking into account that the
specific heat increases in going from the symmetrical to the
asymmetrical AF phase \cite{land1}, we obtain
\begin{equation} \gamma_0(t)=A+\frac{B}{\sqrt{|t-1|}}.\label{TNN}
\end{equation} Here, $B=B_{\pm}$ are the proportionality factors which
are different for the two sides of the phase transition, the
parameters $A=A_{\pm}$ related to the corresponding specific heat
$(C/T)_{\pm}$ are also different for the two sides, and ``$+$''
stands for $t>1$, ``$-$'' stands for $t<1$.

\begin{figure} [! ht]
\begin{center}
\vspace*{-0.8cm}
\includegraphics [width=0.49\textwidth]{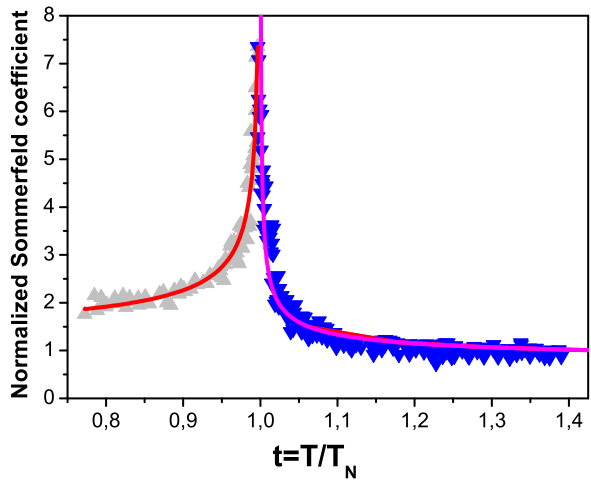}
\end{center}
\vspace*{-0.8cm} \caption{The temperature dependence of the
normalized Sommerfeld coefficient $\gamma_0/A_{+}$ as a function of
the normalized temperature $t=T/T_N(B=0)$ given by the formula
\eqref{TNN} is shown by the solid line. The normalized Sommerfeld
coefficient is extracted from the facts obtained in measurements on
$\rm YbRh_2Si_2$ at the AF phase transition \cite{TNsteg} and shown
by the triangles.}\label{fig3}
\end{figure}

The attempt to fit the available experimental data for
$\gamma_0=C(T)/T$ in $\rm YbRh_2Si_2$ at the AF phase transition in
zero magnetic fields \cite{TNsteg} by the function \eqref{TNN} is
reported in Fig. \ref{fig3}. We show there the normalized
Sommerfeld coefficient $\gamma_0/A_{+}$ as a function of the
normalized temperature $T/T_N(B=0)$. It is seen that the normalized
Sommerfeld coefficient $\gamma_0/A_{+}$ extracted from $C/T$
measurements on $\rm YbRh_2Si_2$ \cite{TNsteg} is well described in
the entire temperature range around the AF phase transition by the
formula \eqref{TNN} with $A_{+}=1$.
\begin{figure} [! ht]
\begin{center}
\vspace*{-0.8cm}
\includegraphics [width=0.49\textwidth]{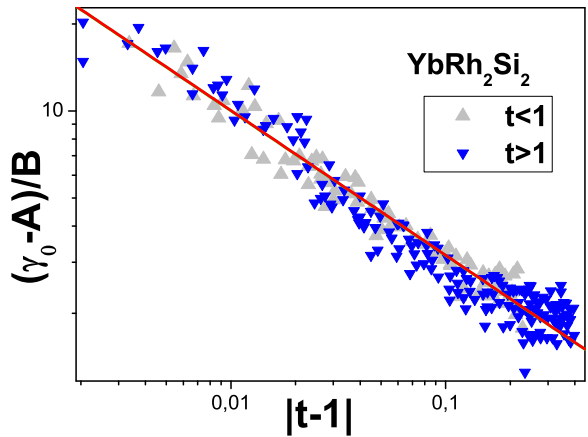}
\end{center}
\vspace*{-0.8cm} \caption{The temperature dependence of the ratios
$(\gamma_0-A)/B$  for $t<1$ and $t>1$ versus $|1-t|$ given by the
formula \eqref{TN1} is shown by the solid line. The ratios are
extracted from the facts obtained in measurements of $\gamma_0$ on
$\rm YbRh_2Si_2$ at the AF phase transition \cite{TNsteg} and
depicted by the triangles as shown in the legend.}\label{fig4}
\end{figure}

Now transform Eq. \eqref{TNN} to the form
\begin{equation} \frac{\gamma_0(t)-A}{B}=\frac{1}{\sqrt{|t-1|}}.\label{TN1}
\end{equation} It follows from Eq. \eqref{TN1} that the
ratios $(\gamma_0-A)/B$  for $t<1$ and $t>1$ versus $|1-t|$ are to
collapse into a single line in logarithmic-logarithmic plot. The
extracted from experimental facts \cite{TNsteg} ratios are depicted
in Fig. \ref{fig4}, coefficients $A$ and $B$ are taken from the
fitting $\gamma_0$ shown in Fig. \ref{fig3}. It is seen from Fig.
\ref{fig4} that the ratios $(\gamma_0-A)/B$ shown by the upward and
downwards triangles for $t<1$ and $t>1$ respectively do collapse
into the single line shown by the solid straight line.

A few remarks are in order here. The good fitting shown in Figs.
\ref{fig3} and \ref{fig4} of the experimental facts by the
functions \eqref{TNN} and \eqref{TN1} with the critical exponent
$\alpha=1/2$ allows us to conclude that the specific-heat
measurements on $\rm YbRh_2Si_2$ \cite{TNsteg} are taken near the
tricritical point and to predict that the second order AF phase
transition in $\rm YbRh_2Si_2$ changes to the first order under the
application of magnetic field as it is shown by the arrow in Fig.
\ref{fig2}. It is seen from Fig. \ref{fig3} that at $t\simeq1$ the
peak is sharp, while one would expect that anomalies in the
specific heat associated with the onset of magnetic order are broad
\cite{TNsteg,PTsteg,lohn}. Such a behavior presents fingerprints
that the phase transition is to be changed to the first order one
at the tricritical point, as it is shown in Fig. \ref{fig2}. As
seen form Fig. \ref{fig3}, the Sommerfeld coefficient is larger
below the phase transition than above it. This fact is in accord
with the Landau theory stating that the specific heat is increased
when passing from $t>1$ to $t<1$ \cite{land1}.

\section{Entropy in $\rm \bf YbRh_2Si_2$ at low temperatures}

It is instructive to analyze the evolution of magnetic entropy in
$\rm YbRh_2Si_2$ at low temperatures. We start with considering the
derivative of magnetic entropy $dS(B,T)/dB$ as a function of
magnetic field $B$ at fixed temperature $T_f$ when the system
transits from the NFL behavior to the LFL one as shown by the
horizontal solid arrow in Fig. \ref{fig2}. Such a behavior is of
great importance since exciting experimental facts \cite{geg1} on
measurements of the magnetic entropy in $\rm YbRh_2Si_2$ allow us
to analyze reliability of the employed theory and to study the
scaling behavior of the entropy when the system in its NFL,
transition and LFL states.
\begin{figure} [! ht]
\begin{center}
\includegraphics [width=0.47\textwidth]{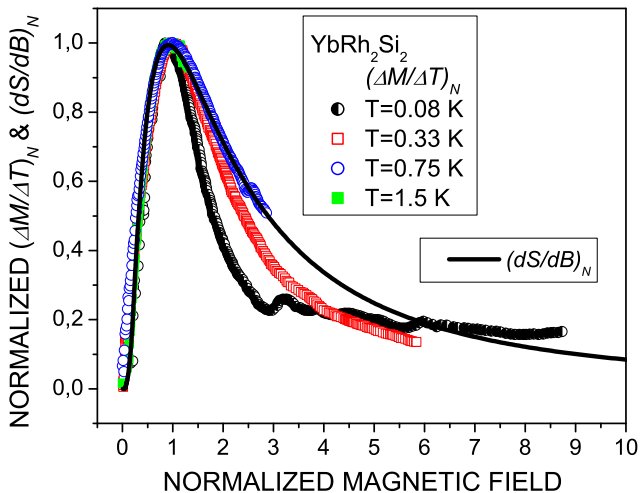}
\vspace*{-0.6cm}
\end{center}
\caption {The behavior of the normalized derivative
$(dS/dB)_N\simeq (\Delta M/\Delta T)_N$ versus normalized magnetic
field when the system transits from the NFL region to the LFL one
along the horizontal solid line shown in Fig. \ref{fig2}.
Normalized magnetization difference divided by temperature
increment $(\Delta M/\Delta T)_N$ versus normalized magnetic field
at fixed temperatures listed in the legend is extracted from the
facts collected  on $\rm YbRh_2Si_2$ \cite{geg1}. Our calculation
of the normalized derivative $(dS/dB)_N$ versus normalized magnetic
field is shown by the solid line.} \label{fig5}
\end{figure}

According to the well-known thermodynamic equality $dM/dT=dS/dB$,
and $\Delta M/\Delta T\simeq dS/dB$. To carry out a quantitative
analysis of the scaling behavior of $dS(B,T)/dB$, we calculate the
entropy $S$ as a function of $B$ at fixed temperature $T_f$ within
the model of homogeneous electron liquid taking into account that
the electronic system of $\rm YbRh_2Si_2$ is located at FCQPT
\cite{plaq}. Fig. \ref{fig5} reports the normalized $(dS/dB)_N$ as
a function of the normalized magnetic field. The normalized
function $(dS/dB)_N$ is obtained by normalizing $(-dS/dB)$ by its
maximum taking place at $B_M$, and the field $B$ is scaled by
$B_M$. The measurements of $-\Delta M/\Delta T$ are normalized in
the same way and depicted in Fig. \ref{fig5} as $(\Delta M/\Delta
T)_N$ versus normalized field. It is seen from Fig. \ref{fig5} that
our calculations shown by the solid line are in good agreement with
the facts and the scaled functions $(\Delta M/\Delta T)_N$
extracted from the facts show the scaling behavior in wide range
variation of the normalized magnetic field $B/B_M$.

Now we are in position to evaluate the entropy $S$ at temperatures
$T\lesssim T^*$ in $\rm YbRh_2Si_2$. At $T<T^*$ the system in its
LFL state, the effective mass is independent of $T$, and is a
function of magnetic field $B$ \cite{obz,geg}
\begin{equation}\label{MB}
\frac{M}{M^*(B)}=a_2\sqrt{\frac{B}{B_{c0}}-1},
\end{equation} where $a_2$ is a parameter.
In the LFL state at $T<T^*$ when the system moves along the
vertical arrow shown in Fig. \ref{fig2}, the entropy is given by
the well-known relation, $S=M^*T\pi^2/p_F^2=\gamma_0T$ \cite{land}.
Taking into account Eqs. \eqref{BC0} and \eqref{MB} we obtain that
at $T\simeq T^*$ the entropy is independent of both magnetic field
and temperature, $S(T^*)\simeq \gamma_0T^*\simeq S_0\simeq
a_1MT_N\pi^2/a_2p_F^2$. Upon using the data \cite{geg}, we obtain
that for fields applied along the hard magnetic direction
$S_0(B_{c0}\,\|\,c)\sim0.03R\ln2$, and for fields applied along the
easy magnetic direction $S_0(B_{c0}\,\bot \,c)\sim0.005R\ln2$.
Thus, in accordance with facts collected on $\rm YbRh_2Si_2$
\cite{geg} we conclude that the entropy contains the
temperature-independent part $S_0$ \cite{obz,yakov} which gives
rise to the tricritical point.\\

\section{Conclusions}

We have predicted that the curve of the second order AF phase
transitions in $\rm YbRh_2Si_2$ passes into the curve of the first
order ones at the tricritical point under the application of
magnetic field. Moreover, we have shown that in the NFL region
formed by FCQPT the curve of any second order phase transition
passes into a curve of the first order one at the tricritical point
leading to the violation of the critical universality of the
fluctuation theory. This change is generated by the
temperature-independent entropy $S_0$ formed behind FCQPT. Near the
tricritical point the Landau theory of second order phase
transitions is applicable and gives the critical index
$\alpha\simeq1/2$. Bearing in mind that a theory is an important
input in understanding of what we observe, we demonstrate that this
value of $\alpha$ is in good agreement with the specific-heat
measurements on $\rm YbRh_2Si_2$ \cite{TNsteg} and conclude that
the critical universality of the fluctuation theory is violated at
the AF phase transition \cite{TNsteg} since the second order phase
transition is about to change to the first order one making
$\alpha\to1/2$.

\section{Acknowledgements}

This work was supported in part by the grants: RFBR No. 09-02-00056
and the Hebrew University Intramural Funds. V.R.S. is grateful to
the Lady Davis Foundation and thanks the generosity of the
Forchheimer Fund for supporting his visit to the Hebrew University
of Jerusalem.

\end{document}